\documentclass[aip,numerical]{revtex4}
\usepackage{graphicx}
\usepackage{fancyhdr}
\usepackage{amsmath} % American Mathematics Society standards
\usepackage{bm}% bold math
\usepackage{amsxtra}
\usepackage{amssymb}
\usepackage{amsthm}
\usepackage{latexsym}
\usepackage{lscape}

\newcommand{\ttbar}{\ensuremath{\mathrm{t}\bar{\mathrm{t}}}~}
\newcommand{\wboson}{\ensuremath{{\mathrm{W}}}}

\begin{document}

\title{Measurement of the top pair invariant mass distribution
at 7 TeV and search for new physics}


\author{S. Rappoccio for the CMS Collaboration}
\affiliation{Johns Hopkins University, Baltimore, MD, USA}

\begin{abstract}
An overview of searches for new physics in the $\ttbar$ sample
from the CMS Collaboration is presented with data collected
at the Large Hadron Collider at $\sqrt{s} = 7$ TeV. There are several
searches presented, including same-sign dilepton signatures,
semileptonic signatures, and all-hadronic signatures, the latter of
which uses advanced jet reconstruction techniques. 
\end{abstract}

\maketitle

\thispagestyle{fancy}


%%%%%%%%%%%%%%%%%%%%%%%%%%%%%%%%%%
\section{Introduction}

A number of scenarios for physics beyond the Standard Model (BSM) feature new
gauge interactions with favorable couplings to the third-generation
quarks (for instance, see References \cite{mssm,nsd,nsd2,nsd3,nsd4,littlehiggs,ed,rs1,rs2}).
These couplings result in new heavy states which could appear as
resonances in top pair production at the LHC.  For example,
Reference~\cite{rs_gluon_1} shows an example of these 
heavy states expressed as Kaluza-Klein gluons with concrete predictions
of cross sections and branching ratios. Also of note are models 
that have recently been proposed to solve the discrepancy in the top
pair production forward-backward asymmetry from the Tevatron
(Reference~\cite{top_AFB1,top_AFB2,top_AFB3,top_AFB4,top_AFB5}).
For instance, one recent model-independent study of the implications
of the forward-backward
asymmetry is presented in Reference~\cite{top_afb_implications},
which argues that a strong enhancement of the production cross
section of $\ttbar$ pairs must be seen at the LHC for invariant
masses above 1.5 or 2.0 TeV/$c^2$ if the deviation is due to
new physics at a heavy mass scale $\Lambda$. 

Searches for new physics in top pair production have been performed
by the Tevatron experiments~\cite{cdftt1, cdftt2, d0tt}.  The Tevatron
measurements provide the most stringent lower mass limits for a narrow
resonance, where a narrow topcolor leptophobic $\ttbar$ resonance is
excluded for masses below about 800~GeV/$c^2$.

The results presented in this paper are based on
data collected by the CMS experiment 
in 2010 and 2011 at a center-of-mass energy of 7~TeV, between 0.036
and 1.1 fb$^{-1}$. Full descriptions of these analyses are presented
in References~\cite{EXO-11-006,TOP-10-007,EXO-11-055,EXO-11-065}.

%%%%%%%%%%%%%%%%%%%%%%%%%%%%%%%%%%
\section{All Hadronic Decay Channel}

The all-hadronic analysis in Reference~\cite{EXO-11-006} utilizes 0.8
fb$^{-1}$ of data, and exploits the highly-boosted nature of the top quarks 
from the high-mass resonances, namely the fact that 
the top quark's decay products often fall inside a single jet. 
If the boost is not too large,
the decay products are distinguishable in this jet (in particular,
the $\mathrm{W}$ boson decay products), and this information can be used
to reduce the large generic QCD dijet production. 
These decay products within the jet are referred to as ``subjets''. 

A {\it top tagging algorithm}~\cite{catop_theory, catop_cms}  is used to identify
merged top jets by analyzing their substructure.  This is accomplished
by examining the clustering sequence of the jets, and the application
of specialized selection criteria.
The behavior of jets from heavy particles such as top quarks is
different from generic QCD jets. For instance, QCD jets
tend to have very few subjets within them, whereas the jets
that originate from hadronic top decays have three or four subjets.
Furthermore, the kinematics of these subjets is different. 
While the subjets of generic QCD jets tend to be close together and 
one often dominates the jet energy (due to gluon emission in the
final state), the top quark decay products  
share the jet energy more equally
and emerge at wider angles. 

The masses
involved in the process (the top mass and the $\mathrm{W}$ mass) also give strong handles for
such discrimination. The mass of a typical QCD jet exhibits a falling spectrum after a quick
initial rise (Reference~\cite{Ellis:2007ib}), whereas the mass of a
fully merged jet from a top quark is very close to the top mass (170-175 GeV/c$^2$).
It is often possible to identify two of the subjets within the top
jet as coming from the decay of the $\wboson$. Similarly to the mass of the entire
jet, for generic QCD jets the mass of this $\wboson$ candidate has a 
falling spectrum in the region of interest, whereas the $\wboson$ decay products 
from the sequential top decay are very close to the $\wboson$ mass.

For situations where the decay products of the top quark are not
contained entirely in one jet, a technique has also been developed to
discriminate against QCD backgrounds using similar techniques as
described for the fully-merged case.  However, for this purpose 
a tool is deployed that is able to handle more general topologies than the
``top jet tagger'' targeted specifically at the
hadronic top decays. 

The {\it jet pruning algorithm}, presented in Reference~\cite{jetpruning1,jetpruning2}, can be
used to identify substructure from general topologies. 
While this tool has been shown to be slightly less performant on fully
boosted top jets than the
targeted top jet tagger, it is of more general utility for
arbitrary topologies. Reference~\cite{boost2010} has detailed algorithmic
comparisons of the taggers for fully boosted top systems, and for instance, 
for the same
efficiency to identify true top quarks, the probability to misidentify
a generic QCD jet with the jet pruning algorithm is larger by 20-30\%
than the targeted top jet tagger.  

On the other hand, the targeted top tagging tool is not 
immediately applicable to moderately boosted
top quark systems where not all of the decay products are merged, 
and as such, the jet pruning tool is used to develop an 
algorithm that identifies boosted
hadronically-decaying $\wboson$-bosons into one jet (referred to as a
$\wboson$ jet). 
In this case, the fact that the decay products from
generic QCD jets are radiated fairly asymmetrically is exploited, whereas the 
decay products from the $\wboson$ are more symmetric
because they arise from a two-body decay of $\wboson$ boson. 
Top quark candidates are then constructed by combining this $\wboson$ jet with another
jet that is close to it, and form a full top-quark candidate.

\begin{figure}
\centering
\includegraphics[width=0.7\textwidth]{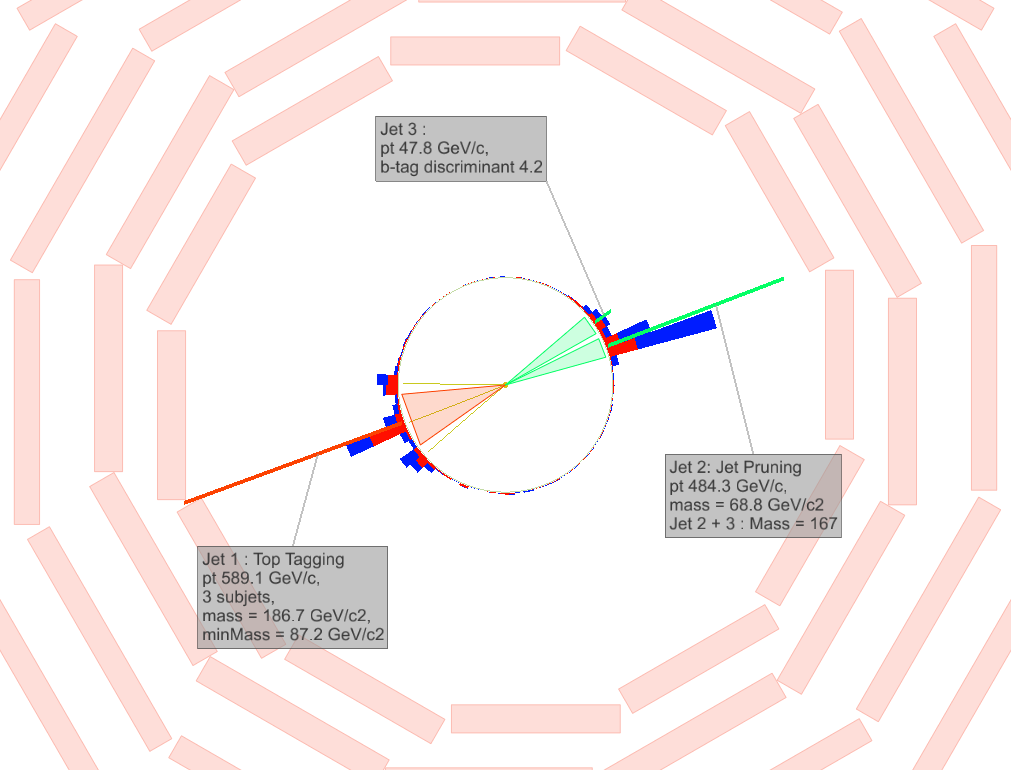}
\caption{Event display of a ``golden'' triply-tagged all-hadronic
  $\ttbar$ candidate. The invariant mass of the $\ttbar$ candidate is 1352.5 GeV/c$^2$.
  This event has a requirement of a top-tagged jet (orange, with
  yellow subjets), a $\wboson$-tagged
  jet and a bottom-tagged jet (both shown in green).
  The electromagnetic
  calorimeter information is shown in red, and the hadronic calorimeter
  information is shown in blue. }
\label{figs:evDisplay_rhophi}
\end{figure}

Figure~\ref{figs:evDisplay_rhophi} shows an event display of
a ``golden'' triply-tagged all-hadronic $\ttbar$ candidate. The event
contains a top-tagged jet, a $\wboson$-tagged jet, and a bottom-tagged jet.
This is a sample that has an enhanced $\ttbar$ contribution. 

\begin{figure}
\centering
\includegraphics[width=0.6\textwidth]{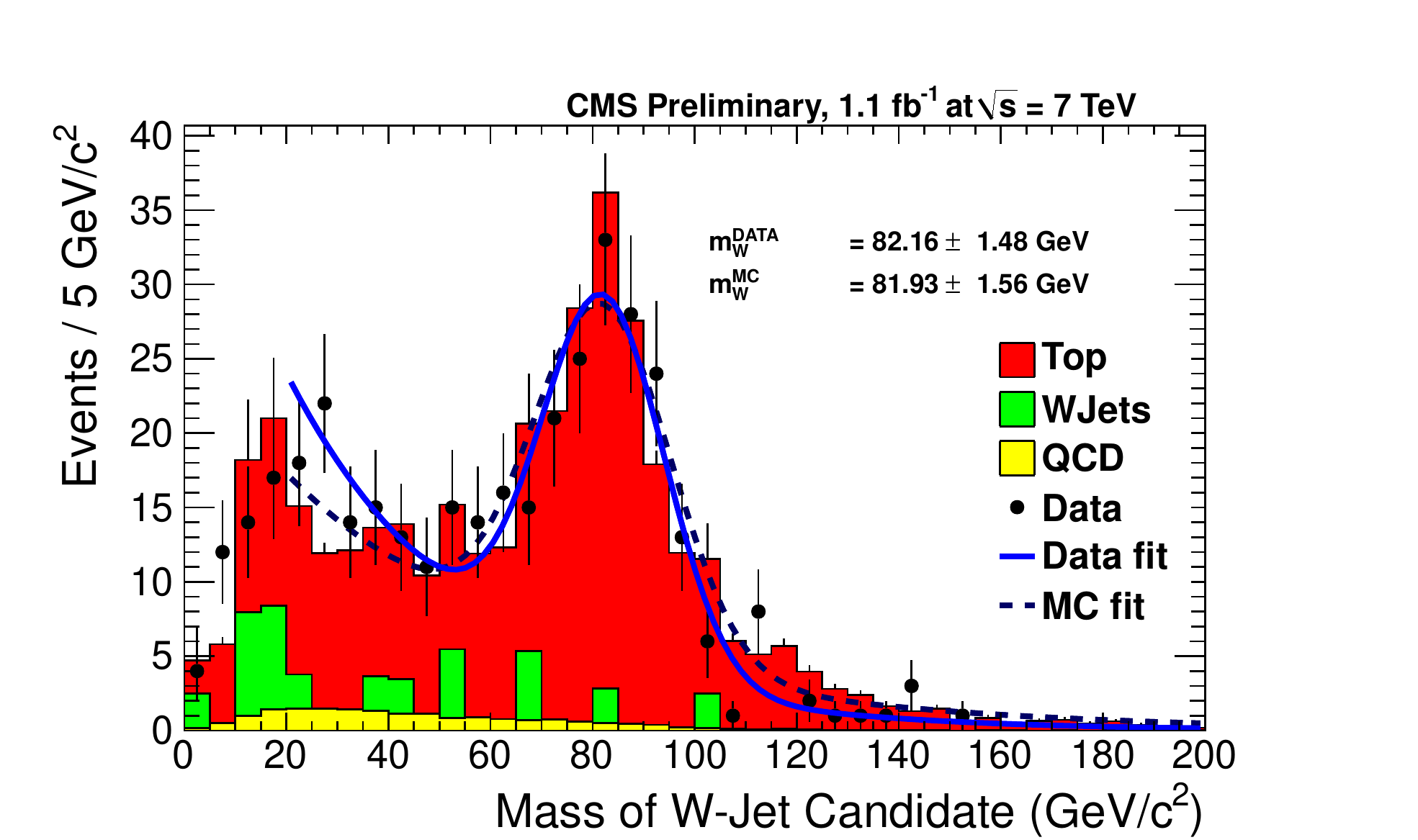}
\caption{Mass of the highest mass jet in a semileptonic
top sample.}
\label{figs:wMass_semileptonic}
\end{figure}

\begin{figure}
\centering
\includegraphics[width=0.6\textwidth]{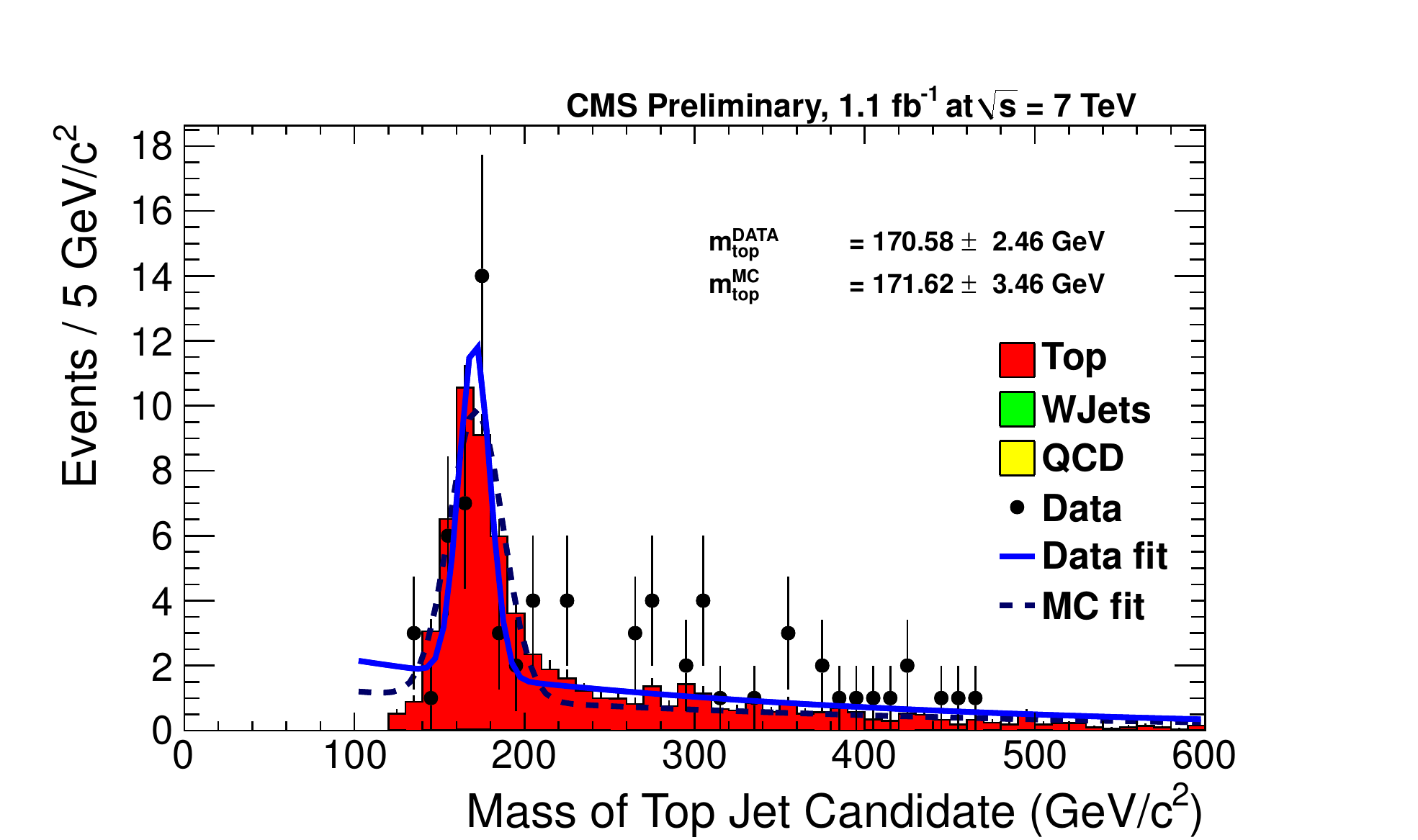}
\caption{Mass of the hadronic top candidate in a semileptonic
top sample.}
\label{figs:topMass_semileptonic}
\end{figure}

A demonstration of the ability to extract boosted jets in a
well-controlled high purity subsample of the data is
shown in Figures~\ref{figs:wMass_semileptonic}
and~\ref{figs:topMass_semileptonic}. In these figures, a sample of
moderately boosted top quarks is extracted from the semileptonic
decay channel by requiring strong transverse momentum cuts to
hemispherically separate the top quark decay products. The
mass of the highest mass jet in the ``hadronic'' hemisphere is shown
(corresponding to the $\wboson$ mass), as
well as the invariant mass of the $\wboson$ candidate with the nearest
jet, which is the mass of the top quark candidate. 
This is a very pure $\ttbar$ sample with negligible backgrounds, and
the data are well-reproduced by the Monte Carlo expectations. The
$\wboson$ peak is quite pronounced, and demonstrates the efficacy of
the methodology in extracting massive boosted hadronic final states
using substructure techniques. 

From this sample, the subjet jet energy scale can be determined from
the difference in the $\wboson$ mass peaks in data and Monte Carlo. 
The subjet energy scale is measured to be 1.01 $\pm$ 0.04 given this
sample. In addition, the efficiency to select $\wboson$ candidate jets
can be measured in the same sample by comparing the event selection
efficiency in data and Monte Carlo, and assuming that the same ratio
(or ``scale factor'') applies to other samples. The
data-to-Monte-Carlo scale factor is determined to be 0.93 $\pm$ 0.13. 

In Reference~\cite{EXO-11-006} there are actually two analyses, one in
the dijet topology (utilizing
two top tags), and one in the trijet topology (utilizing one top tag
and one $\wboson$-tag). The results are combined for a final limit. 

The background estimates for both analyses are taken primarily from data. The small
contribution from Standard Model $\ttbar$ decays is taken from Monte
Carlo, correcting for trigger efficiency, the efficiency of the top-
and $\wboson$-tagging algorithms, and jet energy scale. The largest
background, however, is generic QCD production which has been
mis-identified as having substructure (``mistags''). This background
is estimated by weighting jets in a sample before applying the final
top-jet tag, where the weighting factor is derived from generic dijet
data that has been signal depleted. 

\begin{figure}
\centering
\includegraphics[width=0.6\textwidth]{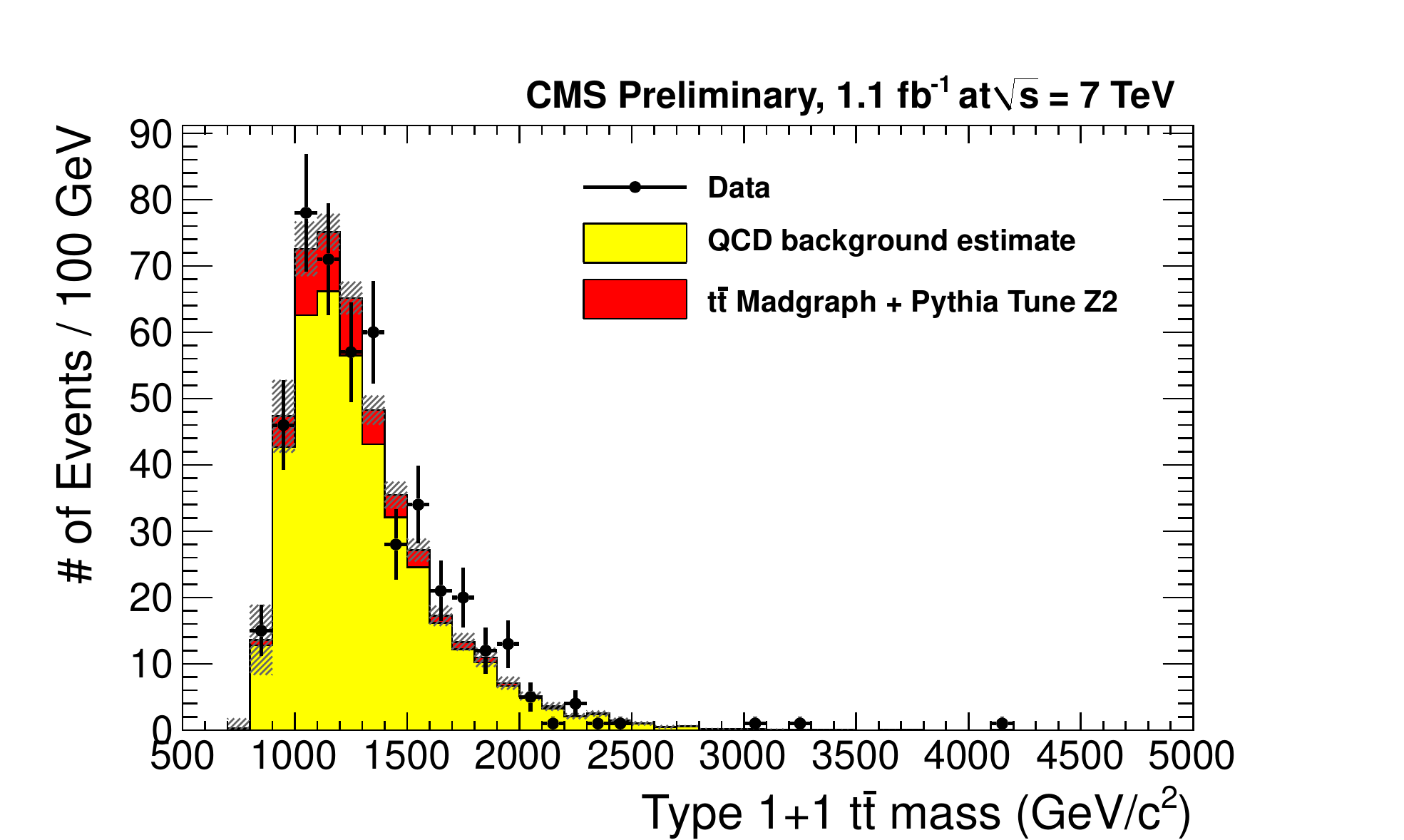}
\caption{Results of the all-hadronic analysis in the dijet topology. 
  The yellow histogram is the QCD estimate from
  the data-driven technique described in the text, and the red histogram is the
  estimate from $\ttbar$ continuum production. 
  The black points are the data.
  The shaded gray boxes
  indicate the statistical and systematic uncertainty on the total background estimate.
   }
\label{figs:ttMassType11}
\end{figure}

\begin{figure}[htbp]
\centering
\includegraphics[width=0.6\textwidth]{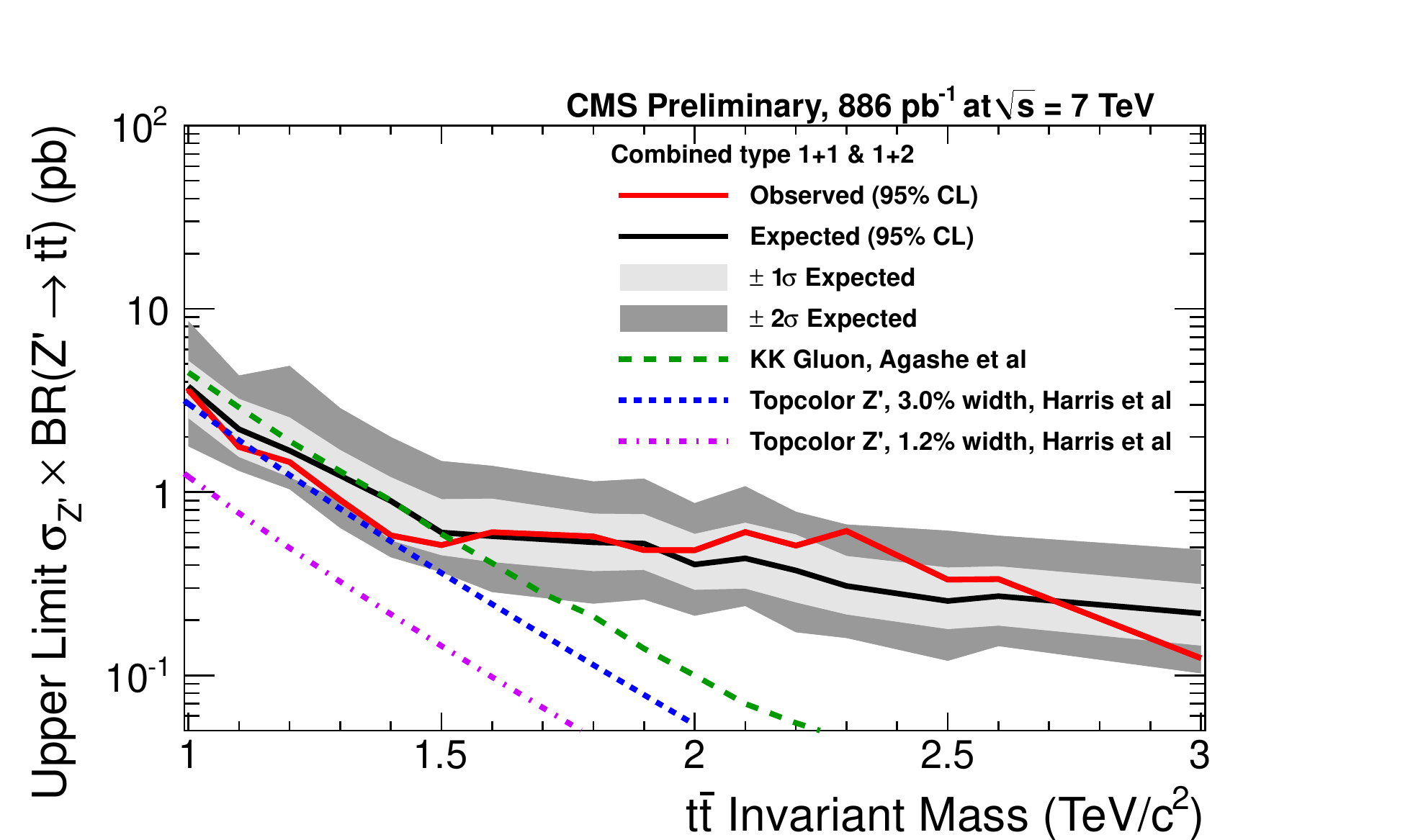}
\caption{The 95\% C.L. upper limit on a product of the production cross
  section of the $\ttbar$ invariant mass and a branching fraction for its decay into $\ttbar$
  pair, as a function of assumed mass. 
  Three theoretical models are examined in shades of purple. From top
  to bottom:
  a Kaluza-Klein gluon from Reference~\cite{rs_gluon_1}, updated to 7 TeV
  via private communication with the authors (Note: the KK gluon model has a width larger than that of the signal Monte Carlo);
  a topcolor $Z'$ model from
  Reference~\cite{hhp} with width 3\%; and 
  a topcolor $Z'$ model from
  Reference~\cite{hhp} with width 1.2\%.}
\label{figs:limit_comb_mcmc}
\end{figure}

Figure~\ref{figs:ttMassType11} shows the results of the event
selection in the dijet topology (similar plots for the trijet topology can
be seen in the original reference). Extremely good agreement is observed
with the prediction, and hence a limit on new physics models is
formulated. The technique chosen is to hypothesize a counting
experiment in a signal window (chosen by the expected size of a narrow
resonance in the $\ttbar$ invariant mass spectrum). A Bayesian
technique is chosen to represent the limits on new physics models,
with Jeffreys priors on the cross section of new physics, and
log-normal priors on the nuisance
parameters. The dijet and trijet topologies are combined in a final
exclusion calculation shown in Figure~\ref{figs:limit_comb_mcmc},
which shows the 68\% and 95\%
credible intervals for observing a resonance at a given mass with a
given cross section times branching ratio. Several theoretical models
are also included for comparison. 

\section{Semileptonic Decay Channel}

There are two analyses from CMS in the semileptonic decay channel. The
first analysis (Reference~\cite{TOP-10-007}, with 0.036 fb$^{-1}$) utilizes standard event
reconstruction techniques assuming that the top quark's decay products
are isotropically distributed (i.e. close to production
threshold). The events are required to have at least three jets,
one well-isolated muon or electron, and significant missing transverse
energy. The events are categorized according to the lepton type (muon
or electron), the number of jets, and
the number of tags in the event, which are then linked by the expected
jet energy scale and bottom-quark tagging efficiency. This technique
is similar to that in the recent measurement of the top pair
production cross section in Reference~\cite{TOP-10-003} (recently
updated to 1 fb$^{-1}$ in Reference~\cite{TOP-11-003}).

The $\ttbar$ candidate is constructed by assigning jets to partons
with a $\chi^2$ sorting method. The combination with the smallest
$\chi^2$ is selected as the best candidate. The top-quark and
$\wboson$-boson masses are used in the constraints of the $\chi^2$
sorting method, and to solve the quadratic ambiguity of the
$z$-component of the neutrino's four-vector. 

The backgrounds are primarily taken from Monte Carlo simulations
comparably to that described in Reference~\cite{TOP-10-003}. The
non-prompt-$\wboson$ backgrounds are taken from data, in sidebands of
the missing transverse energy and isolation selection criteria with a
two-dimensional extrapolation. 

\begin{figure}[htbp]
\centering
\includegraphics[width=0.6\textwidth]{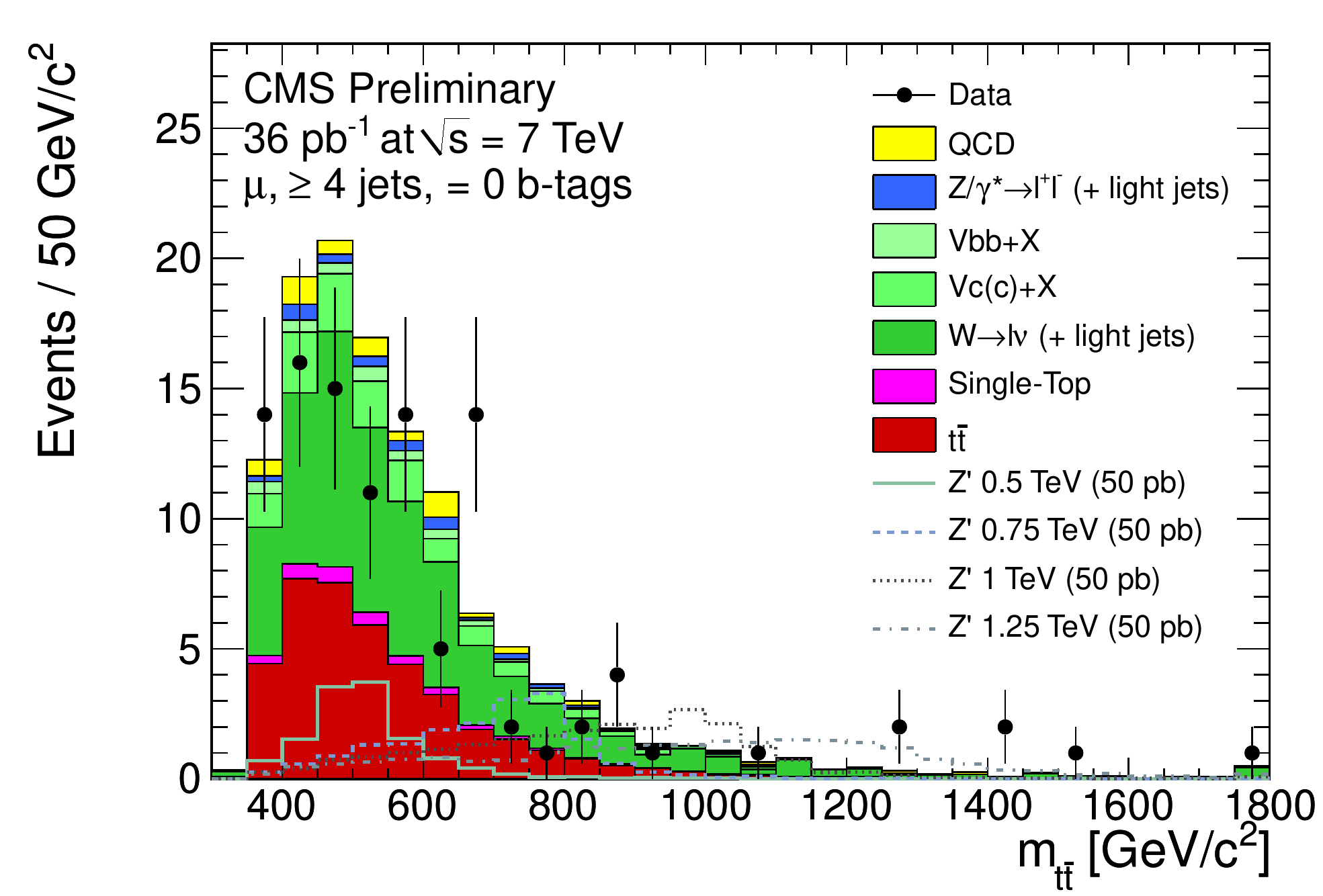}
\caption{Results of event selection for the first semileptonic
  analysis which assumes an isotropically decaying $\ttbar$ candidate.
  The bin shown here is the subsample of data with at least four jets
  and exactly zero bottom-quark-tagged jets.}
\label{figs:mtt_4j_0t}
\end{figure}

\begin{figure}[htbp]
\centering
\includegraphics[width=0.6\textwidth]{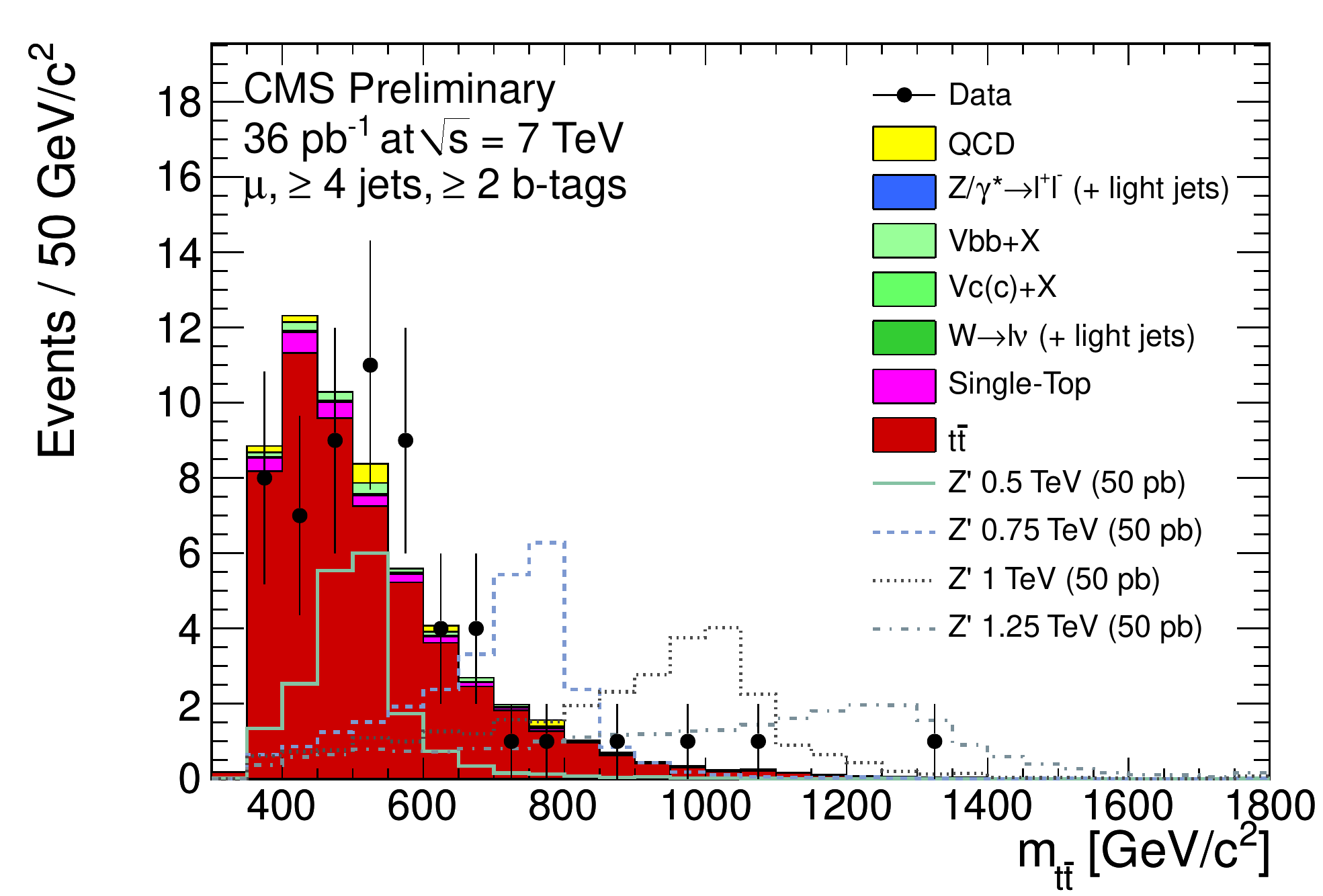}
\caption{Results of event selection for the first semileptonic
  analysis which assumes an isotropically decaying $\ttbar$ candidate.
  The bin shown here is the subsample of data with exactly three jets
  and at least two bottom-quark-tagged jets.}
\label{figs:mtt_3j_2t}
\end{figure}

Figures~\ref{figs:mtt_4j_0t} and~\ref{figs:mtt_3j_2t} show the results
of the event selection for two of the eight subsamples of the
data. Here are shown the muon subsample with at least four jets and exactly
zero bottom-quark-tagged jets, and the muon subsample with exactly three
jets and at least two bottom-quark-tagged jets. Good agreement between
the data and expectation is observed, and hence a limit is set on
possible new physics models. 

\begin{figure}[htbp]
\centering
\includegraphics[width=0.6\textwidth]{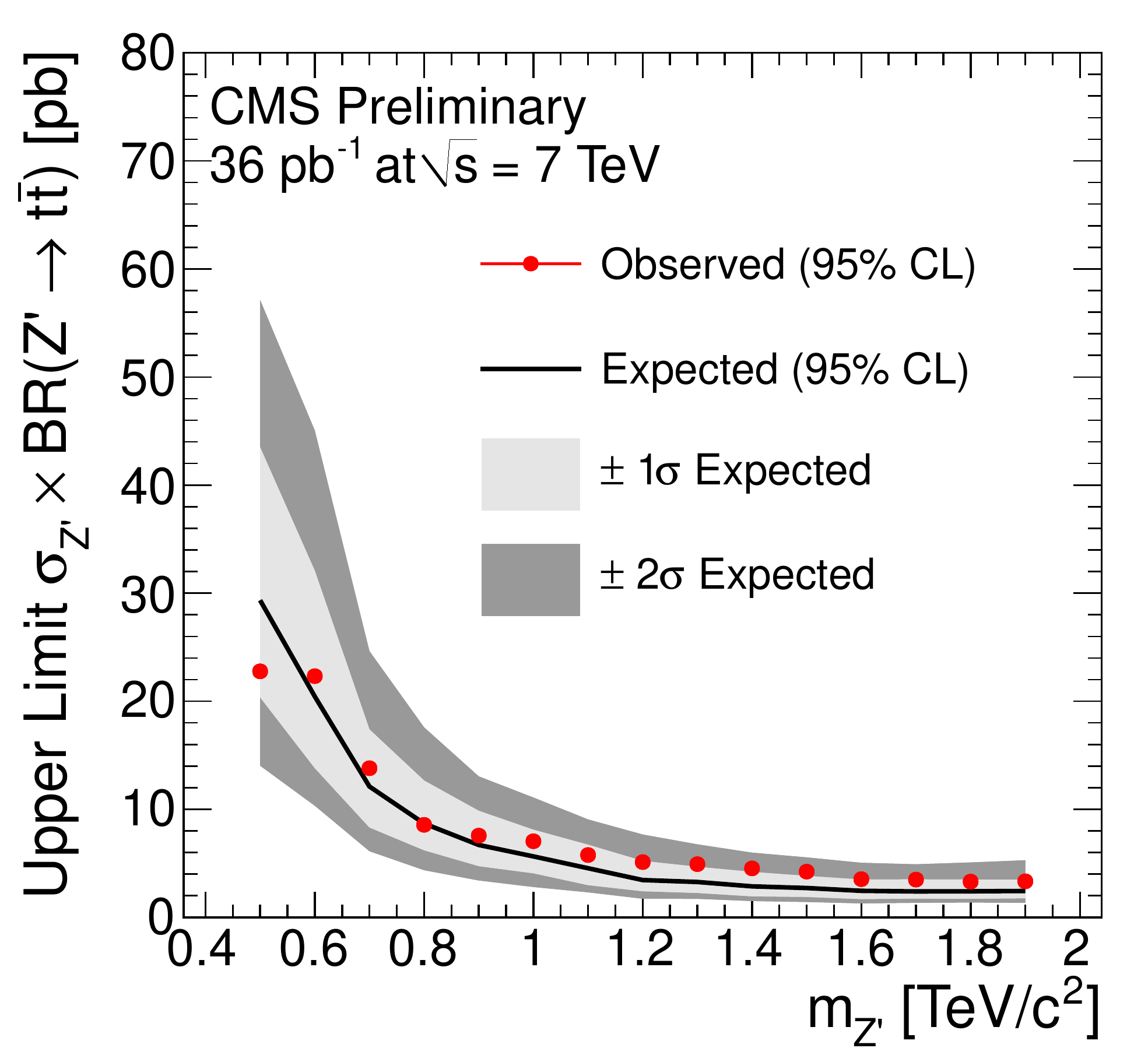}
\caption{The 95\% C.L. upper limit on a product of the production cross
  section of the $\ttbar$ invariant mass and a branching fraction for its decay into $\ttbar$
  pair, as a function of assumed mass. This is for the first
  semileptonic analysis which assumes an isotropically decaying
  $\ttbar$ candidate.}
\label{figs:mtt_lowmass_limits}
\end{figure}

In order to evaluate the statistical limits, a shape analysis is
performed on the $\ttbar$ invariant mass spectrum. A fully Bayesian
approach is taken, as in the all-hadronic case. The same prior
distributions are chosen as well. Figure~\ref{figs:mtt_lowmass_limits}
shows the limits on the production cross section times branching ratio
of a $\ttbar$ resonance at a given mass. Good agreement with the data
is observed.

The second analysis (Reference~\cite{EXO-11-055}, with 1.1 fb$^{-1}$) assumes a boosted
topology and modifies the event selection criteria in order to
efficiently reconstruct the top quarks in this regime. The number of
required jets must be reduced because there is significant jet
merging. The isolation criteria on the muon must also be modified
because the boost of the top quark merges the muon with the nearby
bottom-quark jet. 

Instead of the traditional isolation criterion, a new criterion is
applied in this analysis, which selects events with a two-dimensional
distribution. The first dimension is the angular separation between
the muon and the nearest jet ($\Delta R_{min}$). The second dimension
is the transverse momentum of the leading jet relative to the muon
($p_{T}^{REL}$). Events fail the selection if they have $\Delta R_{min} < 0.5$
and $p_{T}^{REL} < 25$ GeV/c. Furthermore, non-prompt-$\wboson$-boson
backgrounds are suppressed by requiring that the scalar sum of the
lepton transverse momentum and the missing transverse energy
($H_{T,lep}$) be larger than 150 GeV/c$^{2}$. The region with
$H_{T,lep)}$ smaller than 150 GeV/c$^{2}$ is used to normalize the
residual non-prompt-$\wboson$-boson backgrounds. 

\begin{figure}[htbp]
\centering
\includegraphics[width=0.6\textwidth]{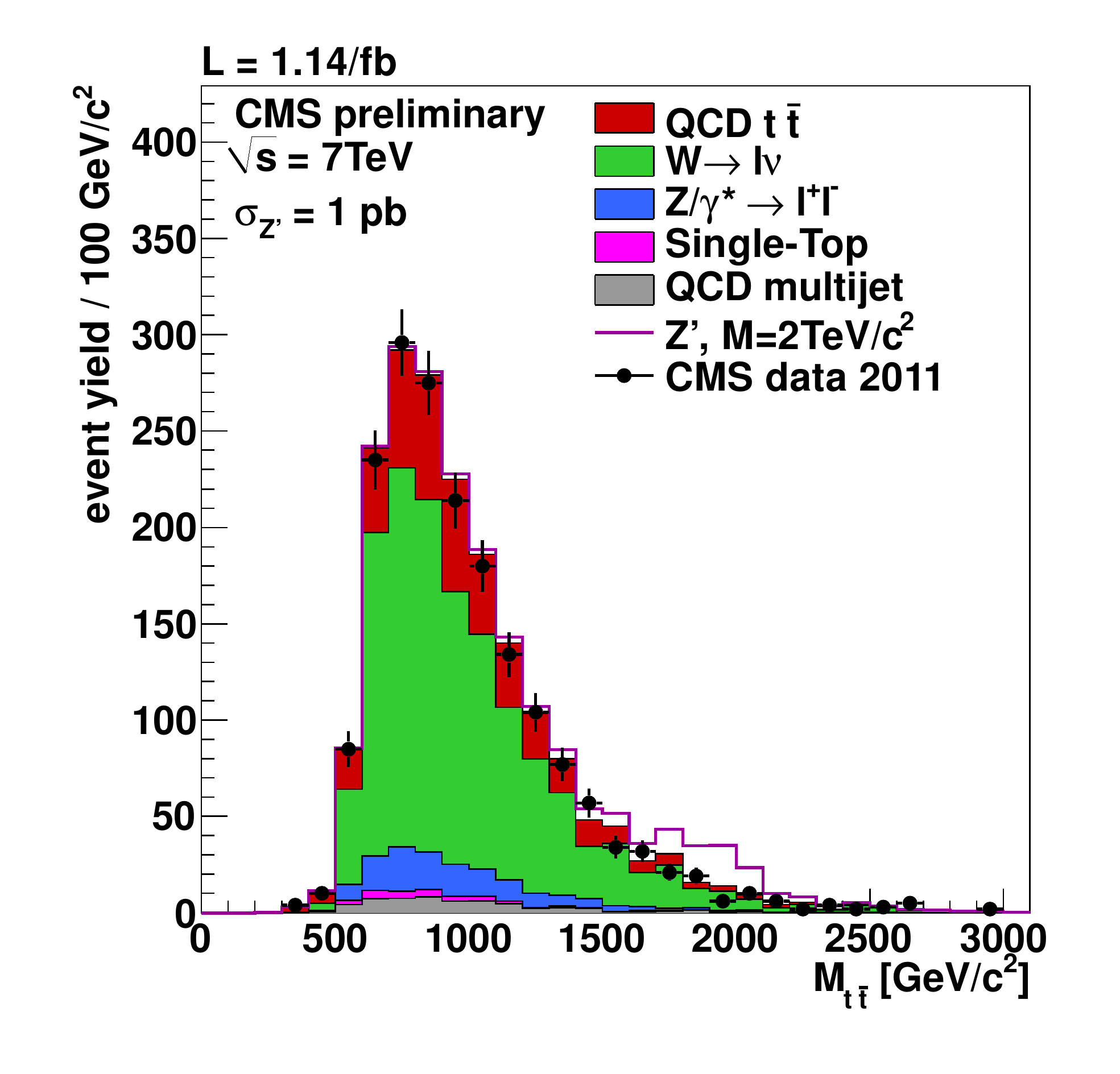}
\caption{Results of event selection for the second semileptonic
  analysis which assumes a boosted $\ttbar$ candidate.}
\label{figs:mtt_highmass}
\end{figure}

\begin{figure}[htbp]
\centering
\includegraphics[width=0.6\textwidth]{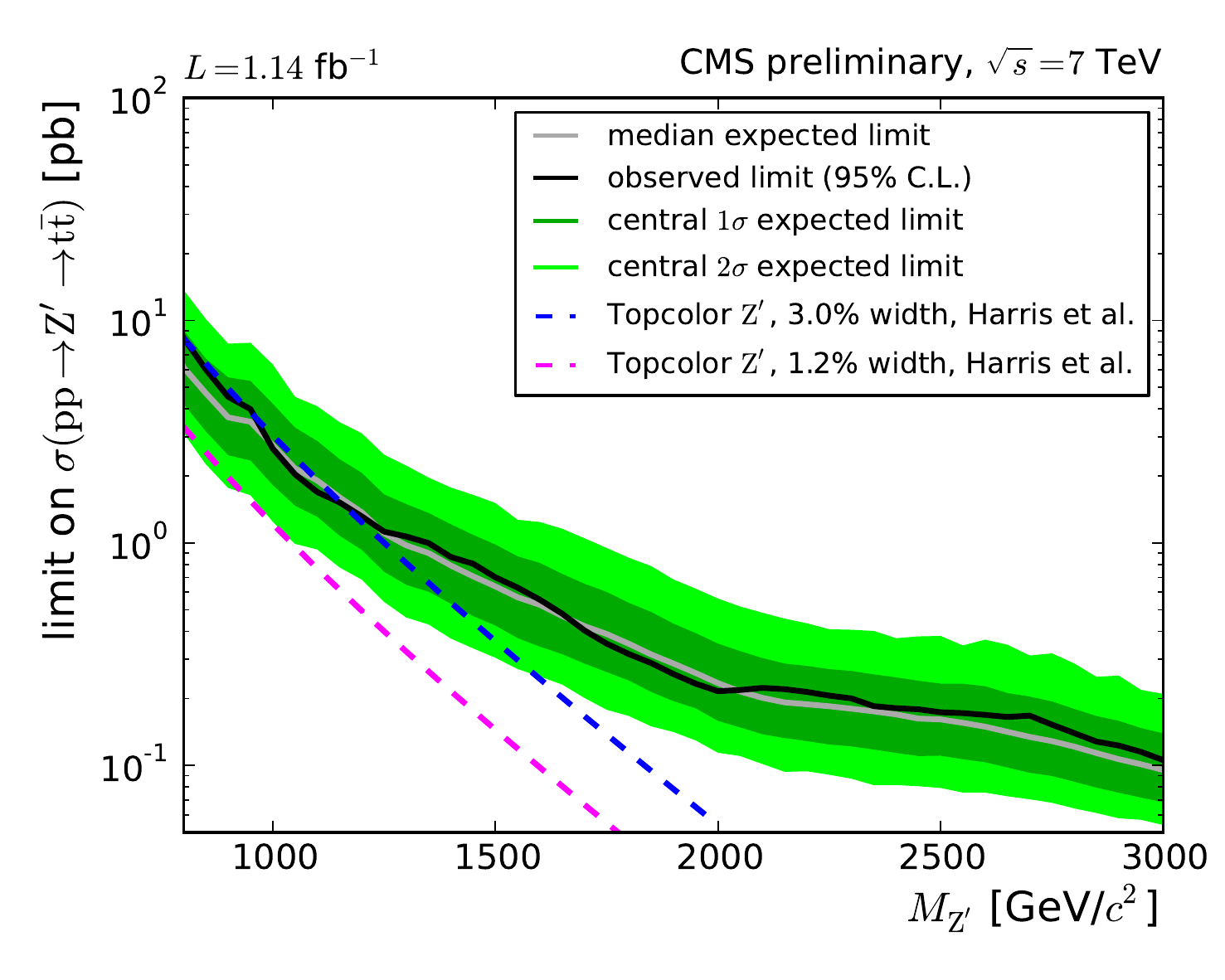}
\caption{The 95\% C.L. upper limit on a product of the production cross
  section of the $\ttbar$ invariant mass and a branching fraction for its decay into $\ttbar$
  pair, as a function of assumed mass. This is for the second
  semileptonic analysis which assumes a boosted 
  $\ttbar$ candidate.}
\label{figs:mtt_highmass_limits}
\end{figure}

The remaining backgrounds are taken from Monte Carlo as in the case of the
first semileptonic analysis. However, in this analysis no
bottom-quark-tagging information is used. 
Figure~\ref{figs:mtt_highmass} shows the results of the event
selection in this boosted analysis, and
Figure~\ref{figs:mtt_highmass_limits} 
shows the 68\% and 95\%
credible intervals for observing a resonance at a given mass with a
given cross section times branching ratio. Several theoretical models
are also included for comparison. The
same mathematical formalism is used as in the first semileptonic
analysis.

\section{Fully Leptonic Decay Channel}

The fully leptonic analysis is a search for anomalous same-sign
dilepton events with 0.036 fb$^{-1}$. 
It is a reinterpretation of a SUSY search
(Reference~\cite{SUSY-10-004}). This search is reinterpreted as a search in
the top sector to examine the hypothesis put forward in
Reference~\cite{berger_topafb} to explain the top forward-backward
asymmetry observed at the Tevatron. In that paper, the
forward-backward asymmetry of top-pair production is enhanced by the
presence of a flavor-changing-neutral-current $Z'$ interaction which
can then produce same-sign top events. 

To test this model, the analysis in the dilepton channel requires two
positive leptons, two or more jets, and missing transverse energy. The
analysis is a counting experiment, with 0.9 $\pm$ 0.6 events expected
from the Standard Model, and 2 events observed. With this data, limits
on the FCNC $Z'$ models can be placed based on the mass of the boson
and the right-handed coupling ($f_R$). 

\begin{figure}[htbp]
\centering
\includegraphics[width=0.6\textwidth]{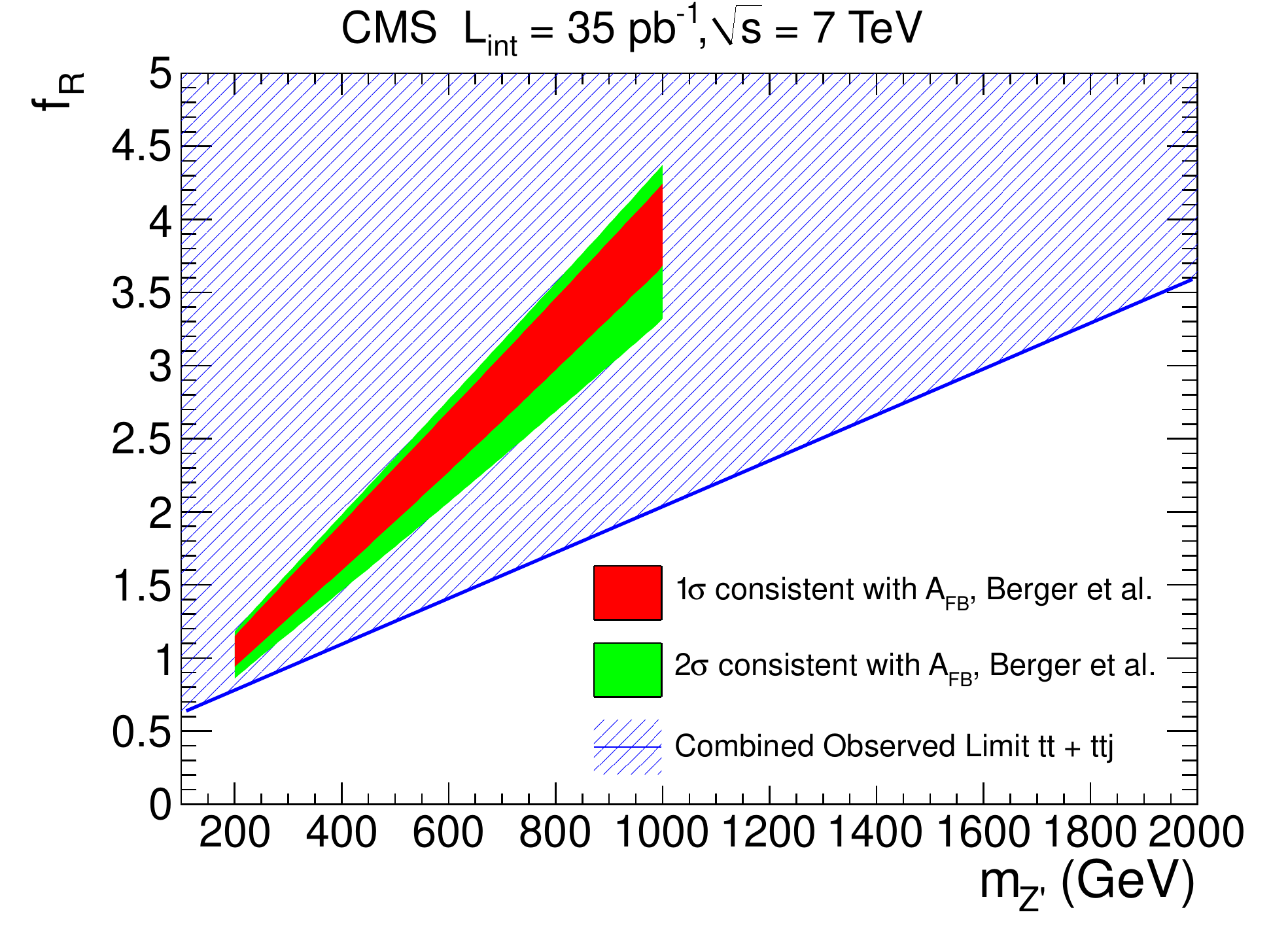}
\caption{The 95\% C.L. upper limit on the model proposed by
  Reference~\cite{berger_topafb} as a function of the boson mass and
  the right-handed coupling parameter $f_R$. The model is disfavored
  by the CMS data. }
\label{figs:top_sscomb}
\end{figure}

Figure~\ref{figs:top_sscomb} shows the 95\% confidence level limit on the model, with
varied boson masses and right-handed couplings. 
The model proposed by
Reference~\cite{berger_topafb} is disfavored.

\subsection{Conclusions}

In conclusion, new physics in the top sector is being rigorously
pursued at CMS with a variety of reconstruction techniques, including
recent advances in boosted jet reconstruction. The plausible new
physics scenarios to explain the top forward-backward asymmetry
anomaly at the Tevatron are beginning to be eliminated, and future
studies will help to further elucidate the situation. 

\bigskip % extra skip inserted
%\begin{thebibliography}{9}   % Use for  1-9  references

%\bibliographystyle{nature}
%\begin{thebibliography}{99} % Use for 10-99 references
\bibliography{rappoccio}

%\end{thebibliography}

\end{document}